\documentclass[prd,tightten,aps,nofootinbib,showpacs,preprintnumbers]{revtex4}
\begin{document}
\title{Confining potential from interacting magnetic and torsion fields}
\author{Patricio Gaete}
\email{patricio.gaete@usm.cl} \affiliation{Departmento de
F\'{\i}sica Universidad T\'ecnica Federico Santa Mar\'{\i}a,
Valpara\'{\i}so, Chile}
\author{Jos\'{e} A. Hela\"{y}el-Neto}
\email{helayel@cbpf.br} \affiliation{Centro Brasileiro de Pesquisas
F\'{\i}sicas, Rua Xavier Sigaud, 150, Urca, 22290-180, Rio de
Janeiro, Brazil}
\date{\today}

\begin{abstract}
Adopting the gauge-invariant but path-dependent variables formalism,
we study the coupling of torsion fields with photons in the presence
of an external background electromagnetic. We explicitly show that,
in the case of a constant electric field strength expectation value,
the static potential remains Coulombic, while in the case of a
constant magnetic field strength expectation value a confining
potential is obtained. This result displays a marked qualitative
departure from the usual coupling of axionlike particles with
photons in the presence of an external magnetic field.
\end{abstract}
\maketitle

\section{Introduction}
One of the most actively pursued areas of research in particle
physics consists of the investigation of extensions of the Standard
Model (SM). This is primarily because the SM does not include a
quantum theory of gravitational interactions. We also recall here
that the SM has many arbitrary parameters, which may seem too many
for a fundamental theory. As is well known, in the search for a more
fundamental theory going beyond the SM string theories \cite{Witten}
are the only known candidate for a consistent, ultraviolet finite
quantum theory of gravity, unifying all fundamental interactions. It
should, however, be noted here that string theories apart from the
metric also predict the existence of a scalar field (dilaton) and an
antisymmetric tensor field of the third rank which is associated
with torsion. This has led to an increasing interest in possible
physical effects of these fields
\cite{Hehl,Audretsch,Rumpf,ShapiroPR}. In addition to the string
interest, torsion fields have also attracted considerable attention
from different viewpoints. Among these, the observed anisotropy of
the cosmological electromagnetic propagation \cite{Nodland,Maroto},
the relativistic and non-relativistic quantum phase acquired by wave
function of a neutral spin-$1/2$ particle with permanent electric
and/or magnetic dipole moment in the presence of a electric and
magnetic fields \cite{Furtado}, and in connection to the interaction
of the light with propagating torsion fields in the presence of an
external magnetic field \cite{Kruglov}. The advent of the CERN Large
Hadronic Collider (LHC) also called attention to test the dynamical
torsion parameters \cite{Belayev} and, related to this issue, the
production of light gravitons \cite{Hagiwara,Ya,Cartier,Jain} at
accelerators justifies the study of dynamical aspects of torsion.

Given the relevance of these studies, it is of interest to improve
our understanding of the physical consequences presented by torsion
fields. Thus, our purpose here is to further explore the impact of
torsion on physical observables, in particular the static potential
between two charges, using the gauge-invariant but path-dependent
variables formalism along the lines of Ref.
\cite{GaeteS,GaeteG,GaeteH}, which is a physically-based alternative
to the usual Wilson loop approach. To this end, we will consider a
system consisting of a gauge field interacting with propagating
torsion fields when there are nontrivial constant expectation values
for the gauge field strength $F_{\mu\nu}$. It is worth recalling at
this stage that the phenomenologically relevant part of the torsion
tensor is dual to a massive axial vector field
\cite{Belayev,ShapiroNPB}, which has a geometric nature. As we shall
see, in the case of a constant electric field strength expectation
value the static potential remains Coulombic. On the other hand, in
the case of a constant magnetic field strength expectation value the
potential energy is linear, that is, the confinement between static
charges is obtained. Incidentally, the above static potential
profile displays a marked departure of a qualitative nature from the
results of axionic electrodynamics \cite{GaeteG}, where the
potential energy is the sum of a Yukawa and a linear potential. One
is thus lead to the interesting conclusion that when torsion fields
are considered the screening part (encoded in the Yukawa potential)
disappears of the static potential profile, describing an exactly
confining phase. In such a case the mass of torsion fields
contribute to the string tension. What this means in physical terms
is that the coupling of torsion fields with photons in the presence
of a constant magnetic field strength expectation, behaves like
small magnetic dipoles in an external magnetic field.

\section{Interaction energy}

As we mentioned above, our immediate objective is to calculate
explicitly the interaction energy between static point-like sources
for the model under consideration. To this end, we will compute the
expectation value of the energy operator $H$ in the physical state
$|\Phi\rangle$ describing the sources, which we will denote by $
{\langle H\rangle}_\Phi$.

The Abelian gauge theory we are considering is governed by the
Lagrangian density \cite{ShapiroNPB,Kruglov}:
\begin{equation}
{\cal L} =  - \frac{1}{4}F_{\mu \nu }^2  - \frac{1}{4}S_{\mu \nu }^2
+ \frac{1}{2}m^2 S_\mu ^2  + \frac{g}{4}S^\lambda  \partial _\lambda
\left( {F_{\mu \nu } \tilde F^{\mu \nu } } \right), \label{Torsion5}
\end{equation}
where $m$ is the mass for the torsion field ($S_\mu$), $S_{\mu \nu }  =
\partial _\mu S_\nu   - \partial _\nu  S_\mu$,
${\widetilde F}_{\mu \nu }  = {\raise0.7ex\hbox{$1$}
\!\mathord{\left/{\vphantom {1 2}}\right.\kern-\nulldelimiterspace}
\!\lower0.7ex\hbox{$2$}}\varepsilon _{\mu \nu \lambda \rho }
F^{\lambda \rho }$, and $g$ is a coupling constant with dimension
$(-2)$ in mass units.

In expression (\ref{Torsion5}) we have used the unique possible form
of the torsion action ( in the low-energy sector) \cite{ShapiroNPB}:
\begin{equation}
{\cal L} = - \frac{1}{4}S_{\mu \nu }^2 + \frac{1}{2}m^2 S_\mu ^2.
\label{Torsion5a}
\end{equation}
It should be noted that the torsion mass term is mandatory, since,
once torsion becomes dynamical, the gravitational excitations
related to the torsion irreducible (irreducibility under Lorentz
group) components become massive, as shown in Ref. \cite{Katanaev,
Katanaev2, ShapiroPR} and \cite{Sezgin}. We would also like to
highlight an important feature of the interacting term that couples
the photon to the pseudo-scalar torsion: $F_{\mu\nu}{\widetilde F}
_{\mu \nu }$, up to a piece that is nothing but the Bianchi identity
for the field-strength tensor, is a measure of the spin density
tensor for the electromagnetic radiation. Then, the photon-torsion
coupling in Eq. (\ref{Torsion5}) may be rewritten as
\begin{equation}
\frac{g}{8}\varepsilon _{\mu \nu \kappa \lambda } \left(
{\partial^\mu \partial ^\rho  S_\rho  } \right)\left( {F^{\nu \kappa
} A^\lambda- F^{\nu \lambda } A^\kappa  } \right), \label{Torsion5b}
 \end{equation}
which renders manifest the coupling of the longitudinal component of
$S^\mu$ to the photon spin density tensor.

Next, by integrating out the $S_\mu$ field in expression
(\ref{Torsion5}), one gets an effective Lagrangian density:
\begin{equation}
{\cal L} =  - \frac{1}{4}F_{\mu \nu }^2  + \frac{{g^2 }}{{8m^2
}}\left( {F_{\mu \nu } \tilde F^{\mu \nu } } \right) \triangle
\left( {F_{\lambda \rho } \tilde F^{\lambda \rho } } \right),
\label{Torsion10}
\end{equation}
where $\triangle\equiv\partial^\mu\partial_\mu$. Now, after splitting
$F_{\mu\nu}$ in the sum of a classical background
$\left\langle {F_{\mu \nu } } \right\rangle$ and a small fluctuation,
$f_{\mu \nu } =\partial _\mu a_\nu -\partial _\nu a_\mu$, the corresponding
Lagrangian density is given by
\begin{equation}
{\cal L}_{eff}  =  - \frac{1}{4}f_{\mu \nu } f^{\mu \nu }  +
\frac{{g^2 }}{{8m^2 }} \left( {v^{\mu \nu } f_{\mu \nu } }
\right)\triangle \left( {v^{\lambda \gamma } f_{\lambda \gamma } }
\right). \label{Torsion15}
\end{equation}
Here, we have simplified our notation by setting $\varepsilon
 ^{\mu \nu \alpha \beta } \left\langle{F_{\mu \nu } } \right\rangle
\equiv v^{\alpha \beta }$ and $\varepsilon ^{\rho \sigma \gamma
 \delta } \left\langle {F_{\rho \sigma } } \right\rangle
 \equiv v^{\gamma \delta }$.
This effective theory thus provides us with a suitable starting
point to study the interaction energy. There is now a non-trivial
point we should raise: the local form of the $4$-photon interaction
Lagrangian of Eq. (\ref{Torsion10}), after the $S^\mu$-field has
been integrated out. According to the Lagrangian (\ref{Torsion5})
and  Eq. (\ref{Torsion5b}), it becomes clear that the transverse
part of $S^\mu$ decouples from the spin density tensor of the
electromagnetic field. So, integrating out the torsion, this
transverse mode does not leave any track. The longitudinal mode
however does couple to the spin density of the photon, as it is
fairly well described in the work of Ref. \cite{Kruglov}. Then, by
virtue of the $\frac{{\partial ^\mu  \partial ^\nu  }}{{m^2 }}$
piece of the $S^\mu$-field propagator, the 4-photon interaction
turns out to be local (upon integration of $S^\mu$), as given above
in Eq. (\ref{Torsion10}).

\subsection{Magnetic case}

We now proceed to obtain the interaction energy in the $v^{0i} \ne
0$ and $v^{ij}=0$ case (referred to as the magnetic one in what
follows). Using this in (\ref{Torsion15}) we then obtain
\begin{equation}
{\cal L}_{eff}  =  - \frac{1}{4}f_{\mu \nu } f^{\mu \nu }  +
\frac{{g^2 }} {{8m^2 }}v^{0i} f_{0i} \Delta v^{0k} f_{0k},
\label{Torsion20}
\end{equation}
where $\mu ,\nu  = 0,1,2,3$ and $i,j,k,l = 1,2,3$. To obtain the
corresponding Hamiltonian, we must carry out the quantization of
this theory. The Hamiltonian analysis starts with the computation of
the canonical momenta $\Pi ^\mu   = f^{\mu 0}  + \frac{{g^2 }}
{{4m^2 }}v^{0\mu } \Delta v^{0k} f_{0k}$, which produces the usual
primary constraint  $\Pi^{0}=0$ while the momenta are $\Pi _i  =
D_{ij} E_j$. Here $E_i  \equiv f_{i0}$ and $D_{ij}  = \delta _{ij}
+ \frac{{g^2 }} {{4m^2 }}v_{i0} \Delta  v_{j0}$. Since ${\bf D}$ is
nonsingular, there exists the inverse ${\bf D}^{-1}$. With this, the
electric field can be written as
\begin{equation}
 E_i  = \frac{1}{{\det D}}\left\{ {\delta _{ij} \det D - \frac{{g^2}}
{{4m^2 }}v_{i0} \Delta v_{j0} } \right\}\Pi _j.  \label{Torsion25}
\end{equation}
 The corresponding canonical Hamiltonian is thus
\begin{equation}
H_C  = \int {d^3 x} \left\{ { - A_0 \partial _i \Pi ^i  - \frac{{M^2
}} {2}\Pi _i \frac{1}{{\left( {\Delta  + M^2 } \right)}}\Pi ^i  +
\frac{{{\bf B}^2 }}{2}} \right\} , \label{Torsion30}
\end{equation}
with
\begin{equation}
M^2  \equiv \frac{{4m^2 }}{{g^2 {\bf v}^2 }} = \frac{{m^2 }}
{{g^2 {\bf {\cal B}}^2 }}. \label{Torsion30a}
\end{equation}
Here, $\bf B$ and ${\bf {\cal B}}$ stand, respectively, for the
fluctuating magnetic field and the classical background magnetic
field around which the $A^\mu$-field fluctuates. $\bf B$ is
associated to the quantum  $A^\mu$-field: $B^i  =
-\frac{1}{2}\varepsilon_{ijk}f^{jk}$, whereas ${\cal B}_i$,
according to our definition for the background  $\left\langle
{F_{\mu \nu } } \right\rangle$ in terms of $v_{\mu \nu }$ is given
by ${\cal B}_i  = \frac{1}{2}v_{0i}$. Time conservation of the
primary constraint yields a secondary constraint. The secondary
constraint is therefore the usual Gauss constraint $\Gamma_1 \left(
x \right) \equiv \partial _i \Pi ^i=0$. Note that the time stability
of this constraint does not induce further constraints.
Consequently, the extended Hamiltonian that generates translations
in time then reads $H = H_C + \int {d^3 }x\left( {c_0 \left( x
\right)\Pi _0 \left( x \right) + c_1 \left( x\right)\Gamma _1 \left(
x \right)} \right)$. Here $c_0 \left( x\right)$ and $c_1 \left( x
\right)$ are arbitrary Lagrange multipliers. It should be noted that
$\dot{A}_0 \left( x \right)= \left[ {A_0 \left( x \right),H} \right]
= c_0 \left( x \right)$, which is an arbitrary function. Since $
\Pi^0 = 0$ always, neither $ A^0 $ nor $ \Pi^0 $ are of interest in
describing the system and may be discarded from the theory. Thus the
Hamiltonian is now given as
\begin{equation}
H = \int {d^3 x} \left\{ {c\left( x \right)\partial _i \Pi ^i -
\frac{{M^2 }}{2}\Pi _i \frac{1}{{\left( {\Delta  + M^2 }
\right)}}\Pi ^i  +  \frac{{{\bf B}^2 }}{2}} \right\}, \label{Torsion35}
\end{equation}
where $c(x) = c_1 (x) - A_0 (x)$.

The quantization of the theory requires the removal of non-physical
variables, which is done by imposing a gauge condition such that the
full set of constraints becomes second class. A particularly
convenient choice is found to be
\begin{equation}
\Gamma _2 \left( x \right) \equiv \int\limits_{C_{\xi x} } {dz^\nu }
A_\nu \left( z \right) \equiv \int\limits_0^1 {d\lambda x^i } A_i
\left( {\lambda x} \right) = 0, \label{Torsion40}
\end{equation}
where  $\lambda$ $(0\leq \lambda\leq1)$ is the parameter describing
the spacelike straight path $ x^i = \xi ^i  + \lambda \left( {x -
\xi } \right)^i $, and $ \xi $ is a fixed point (reference point).
There is no essential loss of generality if we restrict our
considerations to $ \xi ^i=0 $. The choice (\ref{Torsion40}) leads
to the Poincar\'e gauge \cite{GaeteZ,GaeteSPRD}. As a consequence,
the only nonvanishing Dirac bracket for the canonical variables is
given by
\begin{equation}
\left\{ {A_i \left( x \right),\Pi ^j \left( y \right)} \right\}^ *
=\delta{ _i^j} \delta ^{\left( 3 \right)} \left( {x - y} \right) 
- \partial _i^x \int\limits_0^1 {d\lambda x^j } \delta ^{\left( 3
\right)} \left( {\lambda x - y} \right). \label{Torsion45}
\end{equation}

We have finally assembled the tools to determine the interaction
energy for the model under consideration. As mentioned before, in
order to accomplish this purpose we will calculate the expectation
value of the energy operator $H$ in the physical state
$|\Phi\rangle$. Now we recall that the physical state $|\Phi\rangle$
can be written as
\begin{eqnarray}
\left| \Phi  \right\rangle &\equiv& \left| {\overline \Psi  \left(
\bf y \right)\Psi \left( {\bf y}\prime \right)} \right\rangle \nonumber\\
&=&
\overline \psi \left( \bf y \right)\exp \left(
{iq\int\limits_{{\bf y}\prime}^{\bf y} {dz^i } A_i \left( z \right)}
\right)\psi \left({\bf y}\prime \right)\left| 0 \right\rangle,
\label{Torsion50}
\end{eqnarray}
where the line integral is along a spacelike path on a fixed time
slice, and $\left| 0 \right\rangle$ is the physical vacuum state.
The charged matter field together with the electromagnetic cloud (dressing)
which surrounds it, is given by
$\Psi \left( {\bf y} \right) = \exp \left( { - iq\int_{C_{{\bf \xi}
{\bf y}} } {dz^\mu A_\mu  (z)} } \right)\psi ({\bf y})$. Thanks to
our path choice, this physical fermion then becomes $\Psi \left(
{\bf y} \right) = \exp \left( { - iq\int_{\bf 0}^{\bf y} {dz^i  }
A_{i} (z)} \right)\psi ({\bf y})$. In other terms, each of the
states ($\left| \Phi  \right\rangle$) represents a
fermion-antifermion pair surrounded by a cloud of gauge fields to
maintain gauge invariance.

From this and the foregoing Hamiltonian discussion, we then get
\begin{equation}
\Pi _i \left( x \right)\left| {\overline \Psi  \left( \bf y
\right)\Psi \left( {{\bf y}^ \prime  } \right)} \right\rangle  =
\overline \Psi  \left( \bf y \right)\Psi \left( {{\bf y}^ \prime }
\right)\Pi _i \left( x \right)\left| 0 \right\rangle 
+  q\int_ {\bf
y}^{{\bf y}^ \prime  } {dz_i \delta ^{\left( 3 \right)} \left( {\bf
z - \bf x} \right)} \left| \Phi \right\rangle. \label{Torsion55}
\end{equation}
Having made this observation and since the fermions are taken to be
infinitely massive (static) we can substitute $\Delta$ by
$-\nabla^{2}$ in Eq. (\ref{Torsion35}). Therefore, the expectation
value $\left\langle H \right\rangle _\Phi$ is expressed as
\begin{equation}
\left\langle H \right\rangle _\Phi   = \left\langle H \right\rangle
_0 + \left\langle H \right\rangle _\Phi ^{\left( 1 \right)} 
= \left\langle H \right\rangle_0 + \frac{M^2}{2} \left\langle \Phi
\right|\int {d^3 } x\Pi _i \frac{1} {{\left( {\nabla ^2  - M^2 }
\right)}}\Pi ^i \left| \Phi
\right\rangle, \label{Torsion60}
\end{equation}
where $\left\langle H \right\rangle _0  = \left\langle 0
\right|H\left| 0 \right\rangle$.
Using Eq. (\ref{Torsion55}), the $\left\langle H \right\rangle _\Phi ^
{\left( 1 \right)}$ term can be rewritten as
\begin{equation}
\left\langle H \right\rangle _\Phi ^{\left( 1 \right)} = \frac{{M^2q^2
}}{2}\int {d^3 } x\int_{\bf y}^{{\bf y}^{\prime}} {dz_i^{\prime }}
\delta ^{\left( 3 \right)} \left( {{\bf x} - {\bf z}^{\prime}  }
\right) \left( {\nabla ^2 - M^2 } \right)_x^{ - 1}
\int_{\bf y}^{{\bf y}^{\prime}} {dz^i } \delta ^{\left( 3 \right)}
\left( {{\bf x} - {\bf z}} \right).   \label{Torsion70}
\end{equation}
Following our earlier procedure \cite{GaeteS,GaeteS2}, we see that the
potential for two opposite charges located at ${\bf y}$ and ${\bf
y^{\prime}}$ takes the form
\begin{equation}
V = \frac{{q^2}}{{8\pi }} \frac{{m^2 }}{{g^2 {\bf {\cal B}}^2 }}
\mid {\bf y} - {\bf y}^ {\prime} \mid \ln \left( {1 + \frac{{\Lambda
^2 g^2 {\bf {\cal B}}^2 }} {{m^2 }}} \right), \label{Torsion75}
\end{equation}
where $\Lambda$ is a cutoff and $|{\bf y} -{\bf y}^{\prime}|\equiv
L$. Hence we see that the static potential profile displays a
confining behavior. Notice that expression (\ref{Torsion75}) is
spherically symmetric, although the external fields break the
isotropy of the problem in a manifest way. As already pointed out in
our comments soon after the action of Eq. (\ref{Torsion5}), the zero
mass limit is not allowed here, for torsion, from the very
beginning, by virtue of its dynamical character, has to be massive.

Before going ahead, we would like to remark how to give a meaning to
the would-be cutoff $\Lambda$. To do that, we should recall that our
effective model for the electromagnetic field is an effective
description that comes out upon integration over the torsion, whose
excitation is massive. $1/m$, the Compton wavelength of this
excitation, naturally defines a correlation distance. Physics at
distances of the order or lower than 1/m must necessarily take into
account a microscopic description of torsion. This means that, if we
work with energies of the order or higher than m, our effective
description with the integrated effects of $S^\mu$ is no longer
sensible. So, it is legitimate that, for the sake of our analysis,
we identify $\Lambda$ with $m$. Then, with this identification, the
potential of Eq. (\ref{Torsion75})  takes the form below:
\begin{equation}
V = \frac{{q^2 }}{{8\pi }}\frac{{m^2 }}{{g^2 {\cal B}^2 }}\left|
{{\bf y} - {\bf y}^\prime} \right|\ln \left( {1 + g^2 {\cal B}^2 }
\right). \label{Torsion80}
\end{equation}

\subsection{Electric case}

We now want to extend what we have done to the case $v^{0i}=0$ and
$v^{ij}\ne 0$ (referred to as the electric one in what follows). In
such a case the Lagrangian reads
\begin{equation}
{\cal L}_{eff}  = \frac{1}{4}f_{\mu \nu } f^{\mu \nu }  + \frac{{g^2
}} {{8m^2 }}v^{ij} f_{ij} \Delta v^{kl} f_{kl}, \label{Torsion80}
\end{equation}
with $\mu ,\nu  = 0,1,2,3$ and $i,j,k,l = 1,2,3$. Following the same
steps employed for obtaining (\ref{Torsion75}), we now carry out a
Hamiltonian analysis of this model. First, note that the canonical
momenta following from Eq.(\ref{Torsion80}) are $\Pi^\mu=f^{\mu0}$,
which results in the usual primary constraint $\Pi^0=0$ and
$\Pi^i=f^{i0}$. Defining the electric and magnetic fields by $ E^i =
f^{i0}$ and $B^i  = -\frac{1}{2}\varepsilon ^{ijk} f_{jk}$,
respectively, the canonical Hamiltonian takes the form below:
\begin{equation}
H_C  = \int {d^3 x} \left\{ { - A_0 \partial _i \Pi ^i  +
\frac{1}{2}{\bf \Pi} ^2  + \frac{1}{2}{\bf B}^2 } \right\} 
- \frac{{g^2 }}{{8m^2 }}\int {d^3 x} \left\{ {\varepsilon _{ijm}
\varepsilon _{k\ln } v^{ij} B^m \Delta v^{kl} B^n } \right\}.   \label{Torsion85}
\end{equation}
Time conservation of the primary constraint leads to the secondary
constraint, $\Gamma_1(x) \equiv \partial_i\Pi^i=0$, and the time
stability of the secondary constraint does not induce more
constraints, which are first class. It should be noted that the
constrained structure for the gauge field is identical to the usual
Maxwell theory. Thus, the corresponding expectation value is given
by
\begin{equation}
\left\langle H \right\rangle _\Phi   = \frac{1}{2}\left\langle \Phi
\right|\int {d^3 } x\Pi ^2 \left| \Phi  \right\rangle.
\label{Torsion90}
\end{equation}
As we have noted before \cite{GaetePRD}, expression (\ref{Torsion90})
becomes
\begin{equation}
V =  - \frac{{q^2 }}{{4\pi }}\frac{1}{{\mid {\bf y} - {\bf y}^\prime
\mid }}, \label{Torsion95}
\end{equation}
which it is just the Coulomb potential.\\

\section{Final Remarks}

In summary, by using the gauge-invariant but path-dependent
formalism, we have studied the static potential for a gauge theory
which describes the coupling between photons and torsion fields, in
the case when there are nontrivial constant expectation values for
the gauge field strength, $F_{\mu\nu}$. While in the case when
$\langle F_{\mu\nu}\rangle$ is electric-like the static potential
remains Coulombic, we find that, in the case when $\langle
F_{\mu\nu}\rangle$ is magnetic-like, the result is remarkably
different. In fact, when $\langle F_{\mu\nu}\rangle$ is
magnetic-like the potential between static charges displays a
confining behavior. We stress here the role played by the torsion
field in yielding confinement: its mass contribute to the string
tension. Let us also mention here that the singular situation
involving the magnetic field is physically justified: torsion, in
our proposal, couples to the photon spin density tensor, then it
actually probes the magnetic properties of the latter \cite{Perez}
and this suggests us to think that torsion is intrinsically
associated to the magnetic properties (magnetic dipole moment) of
the truly elementary particles. \\

\section{Acknowledgments}

One of us (PG) wants to thank the Field Theory Group of the CBPF for
hospitality and PCI/MCT for support. This work was supported in part
by Fondecyt (Chile) grant 1080260. I. L. Shapiro is kindly acknowledged
for discussions on dynamical torsion.\\

\end{document}